\documentstyle[onecolumn,epsfig]{mn2e}
\def\gtwid{\mathrel{\raise.3ex\hbox{$>$\kern-.75em\lower1ex\hbox{$\sim$}}}}
\def\ltwid{\mathrel{\raise.3ex\hbox{$<$\kern-.75em\lower1ex\hbox{$\sim$}}}}
\def\\{\hfil\break}
\def\ie{{\it i.e.\ }}

\def\etal{{\it et al.\ }}
\def\hmpc{$h^{-1}$Mpc}
\newcommand{\degree}{{\rm o}}
\newcommand{\apj}{ApJ}
\newcommand{\mnras}{MNRAS}

\begin{document}

\title{Bulk Flow and Shear Moments of the SFI++ Survey}
\vskip 0.5cm
\vskip 0.5cm
\author[Feldman \& Watkins]{Hume A. Feldman$^{\star,1}$ \& Richard Watkins$^{\dagger,2}$ \\
$^\star$Department of Physics and Astronomy, University of Kansas, Lawrence, KS 66045.\\
$^\dagger$Department of Physics, Willamette University, Salem, OR 97301\\
emails: $^1$feldman@ku.edu;\, $^2$rwatkins@willamette.edu}

\maketitle

\begin{abstract}

We find the nine bulk--flow and shear moments from the SFI++ survey, as well as for subsamples of  group and field galaxies.
We constrain the velocity power spectrum shape parameter $\Gamma$  in linear theory using these moments. A likelihood function for $\Gamma$ was found after marginalizing over the power spectrum amplitude $\sigma_8\Omega_m^{0.6}$ using constraints obtained from comparisons between redshift surveys and peculiar velocity data.   We have estimated the velocity noise $\sigma_*$  from the data since without it our results may be biased. We also performed a statistical analysis of the difference between the field and group catalogues and found that the results from each reflect the same underlying large scale flows. We found that we can constrain the power spectrum shape parameter to be $\Gamma=0.15^{+0.18}_{-0.08}$ for the groups catalogue and $\Gamma=0.09^{+0.04}_{-0.04}$ for the field galaxy catalogue in fair agreement with the value from WMAP. 
\end{abstract}

\noindent{\it Subject headings}: cosmology: distance scales -- cosmology: large
scale structure of the universe -- cosmology: observation -- cosmology:
theory -- galaxies: kinematics and dynamics -- galaxies: statistics

\section{Introduction}
\label{sec:intro}

Structure formation is assumed to arise from small Gaussian initial fluctuations \cite{BBKS,eh98} amplified by gravitational instability into the large scale structure we observe in current surveys. This scenario has been supported by observations of the bispectrum of galaxy redshift surveys \cite{scoccimarro01,feldman01,verde02} as well as baryonic acoustic oscillations \cite{cole05,eisenstein05}. Since Gaussian random fields are characterized entirely by their power spectrum, on scales where strongly non-linear effects are negligible, the power spectrum provides a glimpse of the primordial fluctuations. On scales $\ltwid100$\hmpc\ (where $h$ is the Hubble constant in units of 100 km s$^{-1}$ Mpc$^{-1}$) the peculiar velocity field of galaxies and clusters is the one of the most promising and reliable probes of the power spectrum \cite{sw95}. Studies of bulk flows of various peculiar velocity surveys  \cite{ph05,pp06,sarkar07} showed the consistency of bulk flow vectors across independent surveys. In a recent paper \cite{wf07} we have shown that there is also a consistency between independent proper distance surveys using both bulk flow and shear moments. 

Recently there are more and more observations of the peculiar velocity field. The surveys are deeper, denser, and more reliable.   As surveys have gotten larger, our understanding of the distance indicators needed to extract the peculiar velocities has also improved. We are getting better at correcting for the various Malmquist biases and other systematic errors, and are able to extract more and better information from surveys \cite{pairwise,radburn04,ph05,sarkar07,wf07}. In this paper we focus on the SFI++ surveys \cite{SFI1,SFI2}, the largest reliable survey of peculiar velocities of the local Universe assembled to date.

Peculiar velocities of individual galaxies or groups are dominated by errors that are proportional to the distances to those objects. 
In addition, individual galaxy or group velocities contain contributions from small scales and are not described well by linear theory.   Previous work has shown that the analysis of individual galaxy velocities using linear theory can lead to biased results \cite{optimal1,optimal2}.   
However, large-scale moments formed from weighted linear combinations of the individual velocities are expected to be better described by linear theory and less susceptible to bias.   Moreover, since large scale moments have much smaller uncertainties than individual velocities, these moments concentrate the information contained in the survey.    In this paper we will work with the nine lowest-order moments corresponding to the three bulk flow and six shear moments, thus keeping much of the information contained in the survey while avoiding possible biases due to small scale flows.   

\section{Analysis}
\label{sec:analysis}

The velocity field can be expressed in a Taylor series
\begin{equation}
{\bf v}_i({\bf r})={\bf u}_i+p_{ij}{\bf r}_j+\dots
\label{eq:velocityfield}
\end{equation}
where ${\bf u}$ is the bulk flow vector and $p_{ij}$ is the shear tensor. The surveys we consider are a set of $N$ galaxies (or groups) each with a position vector ${\bf r}_n$ and a line-of-sight velocity $S_n={\bf v}\cdot{\hat{\bf r}}_n$ with measurement error $\sigma_n$.  Following Kaiser (1991), we create a 9-component vector $g_i({\bf r}) = (\hat r_x,\hat r_y,\hat r_z,r \hat r_x\hat r_x,$ $r\hat r_x\hat r_y,r\hat r_x\hat r_z,r\hat r_y\hat r_y,r\hat r_y\hat r_z,r\hat r_z\hat r_z)$ to model the bulk-flow and shear compontents of the velocity field.   What follows is similar to the development of Watkins \& Feldman (2007); see that paper for details.   

Individual survey velocities can be modeled as
\begin{equation}
S_n =  a_pg_p({\bf r_n}) + \delta_n,
\label{eq:gal-vel}
\end{equation}
where $a_p$ are the moments and $\delta_n$ are Gaussian distributed random variable with  zero mean that accounts for deviations from our model.   We can write the variance of $\delta_n$ as $\sigma_n^2+\sigma_*^2$, where $\sigma_n$ is the measurement error and $\sigma_*$ represents the small-scale higher order linear and non-linear contributions that were not taken into account by our expansion. 

With these assumptions, the maximum likelihood estimate (MLE) for the moments is given by
\begin{equation}
a_p=\sum_{m=1}^Nw_{p,m}S_m\ .
\label{eq:weights}
\end{equation}
with weights given by  \cite{k91,wf07}
\begin{equation}
w_{p,n} = \sum_{q=1}^{ N_{\rm mom}}A_{pq}^{-1} {{\bf g}_q({\bf r}_n)\over  \sigma_n^2+\sigma_*^2}
\label{eq:weights}
\end{equation}
where 
\begin{equation}
A_{pq} = \sum_{m=1}^N {{\bf g}_p({\bf r}_m){\bf g}_q({\bf r}_m) \over  \sigma_m^2+\sigma_*^2}.
\label{eq:aij}
\end{equation}
Here $ N_{\rm mom}=3$ for the bulk flow only and $ N_{\rm mom}=9$ for the bulk flow and and shear moments.  

The covariance matrix of the moments is given by 
\begin{equation}
\langle a_pa_q\rangle = \langle (\sum_n w_{p,n}S_n)(\sum_m w_{q,m}S_m)\rangle 
= \sum_{n,m} w_{p,n} w_{q,m}\langle S_nS_m\rangle.
\label{eq:cov}
\end{equation}
where Eq. (\ref{eq:gal-vel}) can be used to write the covariance matrix for the individual measured velocities $\langle S_nS_m\rangle$  in terms of the velocity field ${\bf v}(\bf{r})$ as
\begin{equation}
\langle S_nS_m\rangle =\langle {\bf \hat r}_n\cdot {\bf v}({\bf r}_n)\ \   {\bf\hat r}_m\cdot {\bf v}({\bf r}_m)\rangle
+ \delta_{nm}(\sigma_*^2 + \sigma_n^2).
\label{eq:galv}
\end{equation}
In linear theory the first term can be expressed as an integral over the density power spectrum $P(k)$, 
\begin{equation}
\langle {\bf \hat r}_n\cdot {\bf v}({\bf r}_n)\ \   {\bf\hat r}_m\cdot {\bf v}({\bf r}_m)\rangle
=  {f^2(\Omega_m)\over 2\pi^2}\int   dk\  W_{mn}(k)\ P(k)
\label{eq:vcov}
\end{equation}
where $f(\Omega_m)\approx\Omega_{m}^{0.6}$ for the scales we are probing and $W_{mn}(k)$ is the angle averaged window function 
 \begin{equation}
W_{mn}(k) = \int {d^2{\hat k}\over 4\pi}\ \left({\bf \hat r}_n\cdot {\bf \hat k}\ \ {\bf \hat r}_m\cdot {\bf \hat k}\right) \exp\left(i{\bf k}\cdot ({\bf r}_n- {\bf r}_m)\right)\ .
\end{equation}

 Plugging Eq. (\ref{eq:galv}) into Eq. (\ref{eq:cov}) and using equation (\ref{eq:vcov}),  the covariance matrix of the moments reduces to two terms, 
 \begin{equation}
 R_{pq} = R^{(v)}_{pq} +  R^{(\epsilon)}_{pq} . 
 \end{equation}
 The first term is given as an integral over the matter fluctuation power spectrum, $P(k)$, 
 \begin{equation} 
 R^{(v)}_{pq}  = {f^2(\Omega_m)\over 2\pi^2}\int_0^\infty
 dk \ \  {\cal W}^2_{pq}(k)P(k),
 \label{eq:covv}
 \end{equation}
where the angle-averaged tensor window function is 
 \begin{equation}
 {\cal W}^2_{pq} (k)=   \sum_{n,m} w_{p,n} w_{q,m}W_{mn}(k)
\label{eq:win}
\end{equation}
 The second term, called the ``noise" term, is given by
 \begin{equation}
 R^{(\epsilon)}_{pq}  = \sum_{n} w_{p,n}w_{q,n}\left( \sigma_n^2 + \sigma_*^2\right)
\label{eq:noise}
  \end{equation}
For the case $a=b$, Eq. (\ref{eq:win}) gives the angle-averaged window function for the moment $a_p$.    This window function tells us which scales contribute to the value of the moment.  For the MLE weights, the bulk flow and shear window functions are determined by the geometry of the survey and the velocity measurement errors.      

Given a peculiar velocity survey and the values of its $ N_{\rm mom}$ moments, we can write the likelihood of a theoretical model used to calculate the covariance matrix as 
\begin{equation}
{\cal L} =  {1\over |R|^{1/2}}\exp\left(-{1\over 2}a_pR^{-1}_{pq}a_q\right).
\label{eq:likelihood}
\end{equation}
We  use this equation in order to place constraints on the parameters of cosmological models, in particular  $\Gamma$, the parameter that determines the shape of the power spectrum as will be discussed below. 

In general, the bulk flow probes scales larger than the diameter of the survey while the shear component probes scales similar to the survey's radius. In this study, we will use a prior constraint on $\sigma_8\Omega_m^{0.6}$ fixes the power spectrum amplitude on scales quite a bit smaller than that of the survey.   Given that the shape parameter $\Gamma$ controls the relative distribution of power between large and small scales (smaller $\Gamma$ corresponds to relatively more power on large scales),  increasing $\Gamma$ generally results in smaller values for the elements of the covariance matrix.    Surveys with large values of bulk flow and shear moments generally favor smaller values of $\Gamma$ (see Eq. \ref{eq:likelihood}).   

How a survey samples the power spectrum is a function of the distribution of the survey objects as well as their measurement errors; both of these properties are reflected in the survey's window function.   The values of the moments are not strictly comparable between surveys since each survey has a unique window function \cite{wf95,sarkar07,wf07}.    However, given the surveys we are considering here, the values of the moments should be highly correlated.   To quantify the agreement, we  use the covariance matrix for the difference between the moments for the two surveys,
\begin{equation}
R^{A-B}_{pq}=  \langle (a_p^A - a_p^B)(a_q^A - a_q^B)\rangle= R^A_{pq} + R^B_{pq}-R^{AB}_{pq}-R^{AB}_{qp},
\label{eq:covdiff}
 \end{equation}
where the cross-terms are given by
\begin{equation} 
 R^{AB}_{pq}  = {\Omega_{m}^{1.2}\over 2\pi^2}\int_0^\infty
 dk \ \ ({\cal W}^{AB})^2_{pq}(k)P(k),
\label{eq:covv-diff}
\end{equation}
and the squared tensor window function is
 \begin{eqnarray}
({\cal W}^{AB})^2_{pq} (k)=
 (A^A)^{-1}_{ps}(A^B)^{-1}_{qt}
 	 \sum_{n,m} {g_s({\bf r}^A_n)g_t({\bf r}^B_m)\over
	((\sigma^A_n)^2+(\sigma^A_*)^2)((\sigma^B_m)^2+(\sigma^B_*)^2)}
	\int {d^2{\hat k}\over 4\pi}\ \left({\bf \hat r}^A_n\cdot {\bf \hat k}\ \ {\bf \hat r}^B_m\cdot {\bf \hat k}\right) 
		\exp\left[i{\bf k}\cdot ({\bf r}^A_n- {\bf r}^B_m)\right]. \nonumber\\
\end{eqnarray}

Here we assume that the nonlinear contributions to the velocities of the galaxies represented by $\sigma_*$ in the two surveys are uncorrelated.    This is not likely to be true in reality, since galaxies in the same local neighborhood are affected by the same small-scale flows.   However, this will always cause us to underestimate the expected amount of correlation between surveys.   Thus our results on how well two surveys agree should be considered as upper bounds.   Given that the covariance matrix is a convolution of the power spectrum and the window function, (Eq.~\ref{eq:covv}) times $f^2(\Omega_m)$, we follow  Watkins \& Feldman (2007) and adopt the Eisenstein \& Hu (1998) parametrizing the transfer function with the parameter $\Gamma$.

\section{Surveys}
\label{surveys}

The surveys we apply our formalisms to are the SFI++ \cite{SFI1} catalogue, as well as subsets of groups and field galaxies. The entire catalogue consists of $\sim$ 5000 spiral galaxies that have I-band Tully-Fisher distances and velocities \cite{TF}. The catalogues represent a homogeneously derived set of positions and peculiar velocities that are corrected for both the homogeneous and inhomogeneous Malmquist biases \cite{SFI2} using the 2MASS redshift survey \cite{2MASS1,2MASS2} for the large-scale structure estimate. The final catalogues we use have 2713 field galaxies with an effective depth of $\sim$75\hmpc\ and 736 groups with an effective depth of $\sim$45\hmpc. These samples are based primarily on the all-sky Spiral Field I band (SFI) and the Spiral Cluster I band (SCI) samples compiled in the 1990's, (see \cite{Haynes1999a,Haynes1999b} and references therein) as well as some data from other samples. The sample covers most of the sky with galactic latitude $b>|10^\degree|$ with some deficiency of galaxies in the declination range of $[-17.5^\degree,-2.5^\degree]$.  The rotation widths for the SFI++ sample come from both the 21 cm line global profile widths ($\sim$60\%) and optical rotation curves ($\sim$40\%).

\section{Results}
\label{results}

In Table 1-a we show the MLE for $a_p$ for the field galaxies (SFI++$_F$), groups (SFI++$_G$), and complete (SFI++) catalogues for the nine bulk flow and shear moments.   The uncertainties given are the diagonal terms in the ``noise" term of the covariance matrix (see Eq. (\ref{eq:noise})). The covariance between the different term, i.e. the off-diagonal terms in the covariance matrix, are at most 10\% of the diagonal terms.   The moments and the value for $\sigma_*$ given for each survey are found using an iterative procedure which maximizes the likelihood; see Watkins \& Feldman (2007) for details.   The likelihood distribution for $\sigma_*$ is quite broad, so that its uncertainty is on order 50\% of its value.   
However, we have found that our results are fairly insensitive to the precise value used in the analysis.

\small
\begin{center}
\begin{tabular}{lcccccccccc}
\multicolumn{9}{c}{\bf Table 1-a Bulk flow and shear magnitude and uncertainties }  \\ 
\multicolumn{9}{c}{Galactic coordinates in the CMB rest frame}\\\hline \hline
name      	 & \multicolumn{3}{c}{Bulk flow} & \multicolumn{6}{c}{Shear}  & $\sigma_*$	\\ 
           	 & \multicolumn{3}{c}{km\ s$^{-1}$} & \multicolumn{6}{c}{km\ s$^{-1}$Mpc$^{-1}$} & km\ s$^{-1}$\\
Components	& x & y & z &  xx & xy & xz & yy & yz & zz & \\ 
\hline \hline
SFI++$_F$& 49$\pm$ 54 & -260$\pm$53 & 100$\pm$45 & 3.2$\pm$1.7 & -3.4$\pm$2.4 & 2.2$\pm$1.8 & 3.7$\pm$1.6 & -2.2$\pm$1.9 & 2.7$\pm$1.4  & 720\\ \hline
SFI++$_G$& 145$\pm$66 & -233$\pm$69 & 71$\pm$47 & 4.3$\pm$2.6 & -3.1$\pm$4.2 & 7.1$\pm$2.9 & -0.43$\pm$2.8 & -5.7$\pm$3.2 & -0.73$\pm$1.9  & 545\\ \hline
SFI++          & 83$\pm$42 & -266$\pm$42 & 73$\pm$33 &  3.3$\pm$1.6 & -3.1$\pm$2.1 & 3.5$\pm$1.6 & 2.8$\pm$1.5 & -3.0$\pm$1.7 & 1.8$\pm$1.4 & 652 \\ \hline
\end{tabular} 
\end{center} 

\normalsize

Table 1-b is the same as Table 1-a with except that the moment values are calculated assuming a model that includes only the three bulk flow moments.  We also show the value for $\sigma_*$ calculated for three degrees of freedom for each survey.    As expected, the $\sigma_*$ values found using the bulk flow model are larger than those found using the bulk flow plus shear model, indicating that at least some of the variation about the bulk flow model is accounted for by shear.   However, overall the values of $\sigma_*$ that we have obtained here are somewhat larger than those found for other surveys (see Watkins \& Feldman 2007), including the original SFI survey \cite{Haynes1999a,Haynes1999b}  whose estimate for $\sigma_*$ was only  $413$ km/s.   This is at least partly due to the fact that the surveys include many objects with very large measurement errors.   Large measurement errors tend to broaden the likelihood distribution of $\sigma_*$ and lead to larger estimates.    This effect is seen in the fact that SFI++$_{\rm G}$, which combines objects into groups and hence has smaller errors than SFI++$_{\rm F}$,  also has a somewhat smaller estimate for $\sigma_*$.   Of perhaps more concern is the fact that the SFI++ survey contains a significant number of galaxies with anomalously large velocities that cannot be accounted for by their measurement errors.    The presence of these galaxies in the survey also results in a larger $\sigma_*$.   We will be exploring these issues further in a future paper.

\begin{center}
\begin{tabular}{lcccc}
\multicolumn{5}{c}{\bf Table 1-b Bulk flow magnitude and uncertainties}  \\ \hline \hline
name      	 & \multicolumn{3}{c}{Bulk flow} & $\sigma_*$ 	\\ 
           	 & \multicolumn{3}{c}{km\ s$^{-1}$} & km\ s$^{-1}$ \\
Components	& x & y & z & \\ \hline \hline
SFI++$_F$&  70 $\pm$ 53 & -281 $\pm$ 52 & 62 $\pm$ 44 & 735 \\ \hline
SFI++$_G$&        91 $\pm$ 63 & -209 $\pm$ 66 & 39 $\pm$ 45 & 554 \\ \hline
SFI++          & 72 $\pm$ 41 & -258 $\pm$ 41 & 50 $\pm$ 32 & 662 \\\hline
\end{tabular} 
\end{center} 
\vspace{0.5cm}

\normalsize
Eq. (\ref{eq:likelihood}) gives the likelihoods for  the parameters of the model for each survey. The theoretical covariance matrix for the velocity moments is completely specified by two parameters; the shape parameter $\Gamma$, and the amplitude $\sigma_8\Omega_m^{0.6}$.    The amplitude parameter has been strongly constrained by comparisons of the velocity field obtained from peculiar velocity data and density fields obtained from redshift surveys. (for a review of these constraints see Pike and Hudson 2005).    We consider the constraint $\sigma_8\Omega_m^{0.6}=0.45\pm 0.05$, which corresponds to that obtained by Zaroubi et al. (2002),  as being most representative of constraints obtained using this method, and we will adopt this constraint as a prior in our calculation of the likelihood function for $\Gamma$.   An estimate of $\sigma_8\Omega_m^{0.6}=0.52\pm 0.06$ obtained directly from the SFI++ survey alone \cite{SFI1} is consistent with this prior.   (An interesting discussion of ways to estimate $\sigma_8$ from peculiar velocity surveys is given by Abate et. al. 2008, see also Watkins et al. (2002) and Feldman et al. 2003a. )   This prior is slightly higher than the value obtained by WMAP three year data \cite{WMAP} but is consistent with it to better than 2$-\sigma$.   By marginalizing over the prior constraint on $\sigma_8\Omega_m^{0.6}$ we are able to calculate the likelihood function for the single parameter $\Gamma$ which determines the shape of the power spectrum (for more details see Watkins \& Feldman 2007).

\begin{figure}
     \includegraphics[width=15cm]{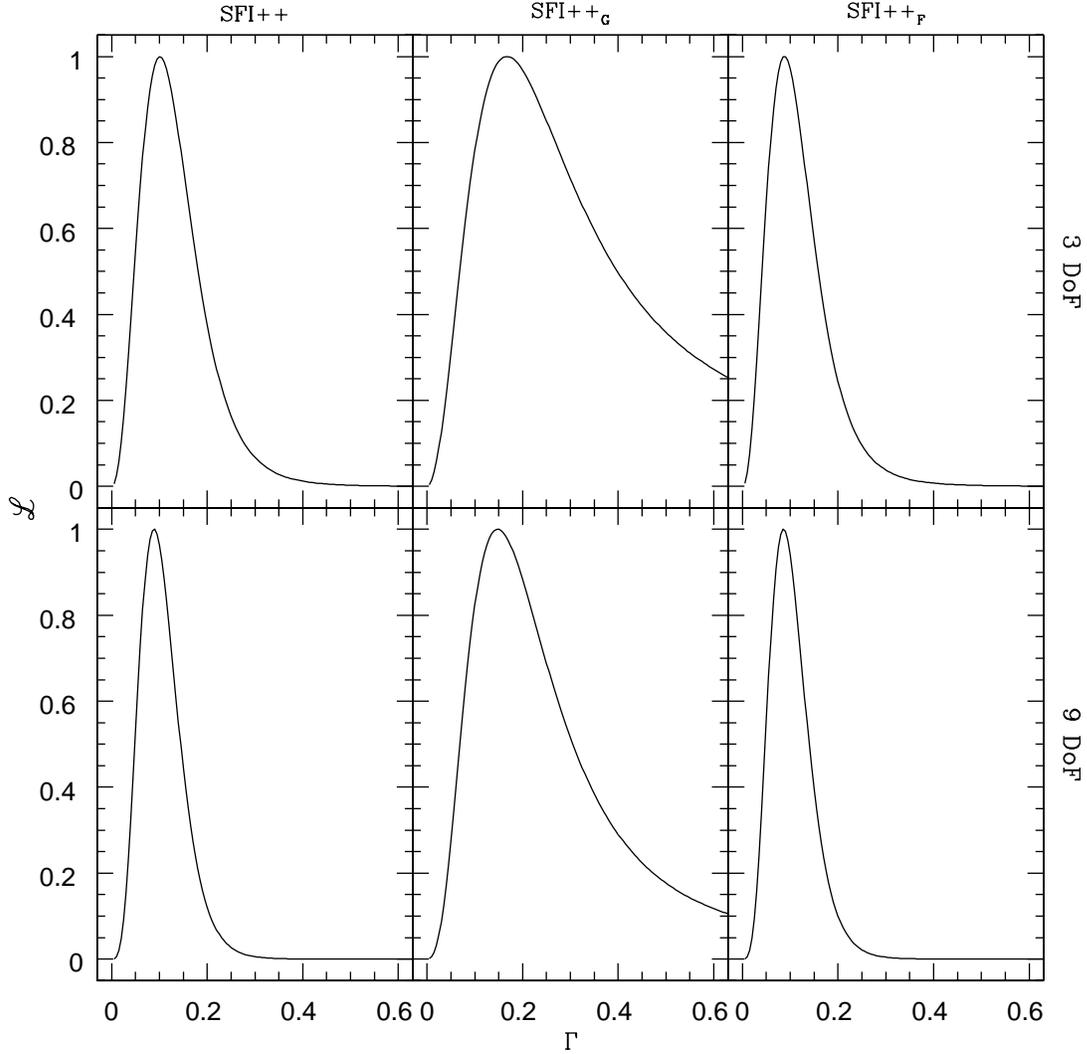}
        \caption{The $\Gamma$ likelihood function for the SFI++ catalogs. The top panels are the maximum likelihood for the bulk-flow moments only, the bottom ones are the likelihoods including all nine moments (bulk-flow and shear). The left panels are for the complete catalog, the middle ones are the likelihoods for the group catalogue, the field galaxies likelihoods are on the right. The inclusion of the shear moments provides tighter constraints on $\Gamma$.}
    \label{fig:likelihood}
\end{figure}

In Fig. \ref{fig:likelihood} we plot the likelihood functions obtained from each of the surveys.   These likelihood functions are asymmetric and have nongaussian tails.   In Table 2 we give the maximum likelihood values of $\Gamma$ for each survey together with the region around the maximum that contains $68\%$ of the probability under the curve.   We also list the $\chi^2$ at the maximum likelihood,  for nine and three degrees of freedom, where 
\begin{equation}
\chi^2 = \sum_{p,q} a_pR^{-1}_{pq}a_q.
\end{equation}
These values show that the peculiar velocity data is consistent with the power spectrum model we are considering.   As can be seen in Table 2 and in Fig. \ref{fig:likelihood}, the maximum likelihood value for $\Gamma$ are consistent for the three and nine DoF, but given the information contained in the six additional moments, the nine DoF likelihood is narrower, providing tighter constrains on $\Gamma$. We would like to stress that the likelihood functions calculated here take into account the entire covariance, including the off diagonal terms.

\vspace{0.5cm}
\begin{center} 
\begin{tabular}{lcccc} 
\multicolumn{5}{c}{\bf Table 2. The maximum likelihood value of $\Gamma$}  \\ 
\multicolumn{5}{c}{\bf and the $\chi^2$ for three and nine degrees of }  \\ 
\multicolumn{5}{c}{\bf freedom for each of the surveys.  The last}  \\ \hline \hline
\multicolumn{5}{c}{\bf entry is the results for the composite survey.}  \\
	        & \multicolumn{2}{c}{3 DoF}   & \multicolumn{2}{c}{9 DoF}\\ \hline
Survey	& $\Gamma$   & $\chi^2$ & $\Gamma$   & $\chi^2$\\ \hline
SFI++$_F$ & $0.09^{+0.06}_{-0.04}$ & 3.13 & $0.09^{+0.04}_{-0.04}$ & 8.63 \\ \hline
SFI++$_G$ & $0.17^{+0.26}_{-0.10}$ & 3.06 & $0.15^{+0.18}_{-0.08}$ & 9.47 \\ \hline
SFI++           & $0.10^{+0.07}_{-0.05}$ & 2.97 & $0.09^{+0.05}_{-0.03}$ & 8.84 \\ \hline
\hline
\end{tabular}

\parbox{3in}{\small }
\end{center} 
\vspace{0.5cm}
\normalsize

Although the group and field galaxy surveys yield consistent results for the value of $\Gamma$ it does not necessarily mean consistency in the actual values of their moments.  In order to check for more detailed compatibility between the surveys we consider the question of whether the differences between the values of the individual  moments of the two surveys, $a_p^F - a_p^G$, (for field and group catalogues, respectively) are consistent with that predicted by the theoretical models, \ie are the measurement errors, the velocity noise, and the differences in how each survey probes the power spectrum large enough to explain the differences in the moments.
To do that we again use a $\chi^2$ analysis;   calculating  the covariance matrix $R^{F-G}_{pq}$ of the difference (Eq.~\ref{eq:covdiff}) we form
\begin{equation}
\chi^2 = \sum_{p,q} (a^F_p-a^G_p)(R^{F-G}_{pq})^{-1}(a^F_q-a^G_q).
\end{equation}
The $\chi^2$ calculated in this way does not depend very strongly on  $\Gamma$ in the region of interest.   For simplicity, then, we report $\chi^2$ values calculated for the single value of $\Gamma = 0.14$. Other values of $\Gamma$ give similar results. Table 3 gives the results of this analysis, which shows good consistency between the catalogues for the favored range of $\Gamma$ values.    Thus the velocity moments of the surveys that we consider agree not only in magnitude, but also in value, inasmuch as they are expected to given measurement errors, velocity noise,  and differences in survey volumes.    

\small
\vspace{0.5cm}
\begin{center} 
\parbox{2.7in}{\small {\bf Table 3:} $\chi^2$ for the differences between the surveys.}
\end{center} 
\begin{center} 
\begin{tabular}{cccccc} 
\hline 
					&& $\chi^2$	&&& $\chi^2$ \\
Surveys				&& 3 DoF 		&&& 9 DoF \\ \hline
SFI++$_F$\ $-$\ SFI++$_G$ 	&&	0.647	&&& 5.59      \\  \hline         
\hline                 
\label{table:diff}
\end{tabular} 
\end{center} 
\vspace{0.5cm}
\normalsize

Since the surveys are consistent with each other, it seems reasonable to combine them into a composite survey which can then be used to obtain the strongest possible constraint on $\Gamma$.    Since the different values of $\sigma_*$ for the various surveys  reflect differences in the populations and distance measures between the surveys, we assign each galaxy in the composite survey the value of $\sigma_*$ of its parent survey.    In Figure \ref{fig:likelihood}, we also show the likelihood function for $\Gamma$ resulting from the composite survey, with the maximum likelihood value being $0.09\pm0.04$ .    In Tables 1-a,b we give the maximum likelihood values for the bulk flow and shear moments for the composite survey.  In Table 2 we present the maximum likelihood value of $\Gamma$ for the composite survey together with its associated $\chi^2$.


In this study we have used a two--parameter model of the power spectrum that is strictly valid only for the zero-baryon case, where theoretically $\Gamma= \Omega_m h$. By interpreting our results for $\Gamma$ it is possible to include the effects of baryons to a first approximation.   Sugiyama (1995) has determined that $\Gamma$ scales with baryonic density $\Omega_b$ as
\begin{equation}
\Gamma = \Omega_m h \exp \left[ -\Omega_b\left(1+\sqrt{2h}/\Omega_m\right)\right].
\label{eq:sug}
\end{equation}
The parameters in this formula are tightly constrained by microwave background studies \cite{WMAP};
specifically, $h=0.732^{+0.031}_{-0.032}$, $\Omega_m=0.241\pm0.034$, and $\Omega_b h^2 = 0.0223^{+0.00075}_{-0.00073}$, which can be combined to give $\Omega_b=0.0416\pm0.0049$. Plugging these values into Eq. (\ref{eq:sug}) and propagating uncertainties gives the result $\Gamma = 0.137\pm0.025$, which is consistent with our results.   While this model is not as accurate as that of Eisenstein \& Hu (1998),  the latter introduces complications in interpretation due to its $\Gamma$ having $k$ dependence.   We have found that the use of the more complicated model in our analysis did not change our results significantly given the precision that can be achieved using the available data.       

The composite SFI++ catalogue gives an estimate of the mean bulk flow to have a magnitude of 288 km s$^{-1}$ $\pm$  71 km s$^{-1}$ toward $l=  285^{\degree}\pm  14^{\degree}$ and $b=  11^{\degree}\pm   10^{\degree}$ where $l$ and $b$ are the galactic longitude and latitude respectively. This result agrees well with other estimates from the SFI catalogue, such as the one obtained by \cite{giovanelli98} to have a magnitude of 200 km s$^{-1}$ $\pm$  65 km s$^{-1}$ toward $l=  295^{\degree}\pm  20^{\degree}$ and $b=  25^{\degree}\pm 20^{\degree}$. It also agrees with the results in \cite{sarkar07} of 330 km s$^{-1}$ $\pm$ 101 km s$^{-1}$ toward $l= 234^{\degree}\pm 11^{\degree}$ and $b=12^{\degree}\pm 9^{\degree}$. 

The SFI++ catalogues discussed in this paper are highly consistent with each other, and provide a strong constraint of the power spectrum shape $\Gamma$. Since the groups catalogue is shallower, it appears that the flows reflected by the bulk flow and shear moments are smaller than that for the field galaxy catalogue. This results in a flatter power spectrum and thus larger MLE for $\Gamma$. This seems to suggest that we have not reached convergence of the velocity field with current generation of surveys. However, this study supports and strengthens the notion that velocity fields have come of age and provide an excellent dynamic probe of the large scale structure of the Universe. The results from this and similar analyses show remarkable agreement between independent catalogues that survey various morphologies, selection criteria, geometries, in galaxy clusters, groups of galaxies and individual galaxy catalogues. Both our surveying abilities and our analyses tools have matured to the point that we can study flows on scales unreachable just a few years ago and provide uniquely dynamical probe to the study of the large--scale structure of the Universe.

\noindent{\bf Acknowlegements:} 
HAF has been supported in part by a grant from the Research Corporation and by the University of Kansas General Research Fund (KUGRF).

\end{document}